\documentclass[reqno,12pt]{article}

\usepackage{graphicx}
\usepackage{amsmath,amsthm,amssymb}

\chardef\bslash=`\\ 

\theoremstyle{definition}

\theoremstyle{remark}

\newcommand{\eval}[2][\right]{\relax
 \ifx#1\right\relax \left.\fi#2#1\rvert}

\begin{document}
\title{\bf{A note on the non-existence of $\sigma$-model solitons in the $2+1$ dimensional AdS gravity}}

\author{Piotr Bizo\'n\footnotemark[1]{} \;  and Arthur Wasserman\footnotemark[2]{}\\
  \footnotemark[1]{} \small{\textit{Institute of Physics,
   Jagellonian University, Krak\'ow, Poland}}\\
   \footnotemark[3]{} \small{\textit{Department of Mathematics,
  University of Michigan, Ann Arbor, Michigan}}}

%
\maketitle
\begin{abstract}
\noindent We show that the gravitating static soliton in the $2+1$
dimensional $O(3)$ $\sigma$ model does not exist in the presence
of  a negative cosmological constant.
\end{abstract}

\section{Introduction}
The $2+1$ dimensional $O(3)$ $\sigma$ model coupled to gravity is
a wave map $X: M \rightarrow N$  from a $2+1$ dimensional
spacetime $(M,g_{ab})$ into a two-sphere $S^2$ with the round
metric $G_{AB}$ defined by the action
\begin{equation}
S = \int_M \left(\frac{R+2\Lambda}{16 \pi G}  + L_{WM}\right) dv_g
\end{equation}
with the Lagrangian density
\begin{equation}
  L_{WM} = -\frac{f^2_{\pi}}{2} g^{ab} \partial_a X^A \partial_b X^B
  G_{AB}.
\end{equation}
Here $\Lambda$ is a cosmological constant, $G$ is the Newton
constant and $f^2_{\pi}$ is the wave map coupling constant. The
product  $\alpha=8\pi G f^2_{\pi}$ is dimensionless. The field
equations derived from (1) are the wave map equation
\begin{equation}
\square_g X^A + \Gamma_{BC}^A(X) \partial_aX^B \partial_bX^C
g^{ab}=0,
\end{equation}
where $\Gamma_{BC}^A(X)$ are the Christoffel symbols of the target
metric $G_{AB}$ and $\square_g$ is the wave operator associated
with the metric $g_{ab}$, and  the Einstein equations
$R_{ab}-\frac{1}{2} g_{ab} R + \Lambda g_{ab} = 8 \pi G T_{ab}$
with the stress-energy tensor
\begin{equation}
  T_{ab} = f^2_{\pi} \left(\partial_a X^A \partial_b X^B
  -\frac{1}{2} g_{ab}( g^{cd} \partial_c X^A \partial_d X^B)\right)
  G_{AB}.
\end{equation}
 In  polar coordinates $X^A=(F,\Phi)$ on the target $S^2$ the metric
 takes the form
\begin{equation}
G_{AB} dX^A dX^B = dF^2 + \sin^2{\!F} \: d\Phi^2.
\end{equation}
For the domain manifold we take a spherically symmetric $2+1$
dimensional spacetime and parametrize the metric using areal
coordinates
\begin{equation}
g_{ab}dx^a dx^b= -e^{-2 \delta} A\: dt^2 + A^{-1} dr^2 + r^2 \:
d\phi^2,
\end{equation}
 where $\delta$ and $A$ are functions of $(t,r)$.
 Next, we assume that the wave maps are corotational, that is
\begin{equation}
F= F(t,r),\quad \Phi=\phi.
 \end{equation}
Equation (3) reduces then to the single semilinear wave equation
(hereafter, primes and dots denote derivatives with respect to $r$
and $t$, respectively)
\begin{equation}
-e^{\delta} (e^{\delta} A^{-1} \dot F)^{\dot{}}
+\frac{e^{\delta}}{r} (r e^{-\delta} A\: F')'=\frac{\sin(2 F)}{2
r^2},
\end{equation}
and the Einstein equations become
\begin{eqnarray}
\dot A &=& - \alpha\: r A \dot F F',
\\
 \delta' &=& -2 \Lambda r -\alpha\: r \left(F'^2 +
A^{-2} e^{2\delta} {\dot F}^2 \right),
\\
A' &=&  - \alpha\: r \left( A F'^2 + A^{-1} e^{2\delta} {\dot F}^2
+ 2 \:\frac{\sin^2{\!F}}{r^2}\right).
\end{eqnarray}
The studies of the initial value problem for this system in the
case of zero cosmological constant, performed first in the flat
spacetime ($\alpha=0$) \cite{bct} and recently also for $\alpha>0$
\cite{cs}, showed that the scale-free static solution plays a
crucial role in the process of singularity formation, namely
singularities form via a static solution shrinking adiabatically
to zero size. In fact, Struwe showed that for equivariant wave
maps in the flat spacetime singularities must form in this way
\cite{s}, in other words non-existence of a nontrivial static
solution implies global regularity. Thus, it seems interesting to
see how the inclusion of a negative cosmological constant affects
the structure of static solutions.

\section{Static solutions for $\Lambda=0$}
Before looking at the static solutions of equations (8)-(11) with
$\Lambda<0$, in this section we review some well known facts about
static solutions for $\Lambda=0$. We first consider the case
$\alpha=0$ which corresponds to the flat spacetime $A=1, \delta=0$
so equation (8) reduces to
\begin{equation}\label{static}
\frac{1}{r} (r F')' = \frac{\sin(2 F)}{2 r^2}.
\end{equation}
The trivial constant solutions of (\ref{static}) are $F=0$ and
$F=\pi$; geometrically these are maps into the north and the south
pole of $S^2$, respectively. The energy of these maps
\begin{equation}\label{energy}
E(F)=\pi \int\limits_0^{\infty} (F'^2+
  \frac{\sin^2\!{F}}{r^2}) \: r dr
\end{equation}
attains the global minimum $E=0$. Note that the requirement that
energy be finite imposes a boundary condition at spatial infinity
$F(\infty)=k \pi$ ($k=0,1,\dots$) which compactifies
$\mathbb{R}^2$ into $S^2$ and thereby breaks the phase space into
infinitely many disconnected topological sectors labelled by the
degree $k$ of the map $S^2\rightarrow S^2$.

The fact that equation (\ref{static}) is scale invariant does not
exclude nontrivial regular solutions with finite energy (Derrick's
argument is not applicable) and, in fact, such solutions are
well-known both in the mathematical literature as harmonic maps
from $\mathbb{R}^2$ into $S^2$ and in the physics literature as
instantons in the two-dimensional euclidean sigma model. One way
to derive them is to use the Bogomol'nyi  identity
\begin{equation}\label{bogom}
E(F)=\pi\int\limits_0^{\infty} ({F'}^2+
  \frac{\sin^2\!{F}}{r^2}) \: r dr = \pi \int\limits_{0}^{\infty}
  \left(\sqrt{r} F'- \frac{\sin\!{F}}{\sqrt{r}}\right)^2 \: dr - 2\pi \cos{F}
   \Bigr\rvert^{\infty}_{0}.
\end{equation}
It follows from (\ref{bogom}) that in the topological sector $k=1$
the energy attains the minimum, $E=4\pi$, on the solution of the
first order equation $ r F'=\sin\!{F}$, which is
\begin{equation}\label{Fs}
F_S(r)=2 \arctan(r/\lambda),
\end{equation}
 where $\lambda$ is a nonzero constant. This solution is a
well-known harmonic map from $R^2$ to $S^2$. We remark in passing
that  this solution can be alternatively obtained in an elegant
geometric way  by composing the identity map between two-spheres
with the inverse of stereographic projection

It has been known for long that the solution (\ref{Fs})  persists
if one couples gravity with zero cosmological constant \cite{cg}.
To see this let us consider equations (8)-(11) and assume that the
fields are time independent. We get
\begin{equation}
\frac{1}{r} e^{\delta} \left( A e^{-\delta} r F'\right)' =
\frac{\sin(2 F)}{2 r^2},
\end{equation}
and (assuming that $\Lambda=0$)
\begin{eqnarray}
 \delta' &=& -\alpha\: r F'^2,\\
A' & = & -\alpha r \left( A F'^2 + \frac{\sin^2{F}}{r^2}\right).
\end{eqnarray}
For regular solutions the boundary conditions at $r=0$ are
\begin{equation}
A(0) =1, \quad \delta(0)=0, \quad F(0)=0, \quad F'(0)=b,
\end{equation}
where $b$ is a free parameter. We want a finite energy degree-one
solution so we require that $A(r)$ and $\delta(r)$ tend to
constants at infinity and $F(\infty)=\pi$ . Such a solution can be
found explicitly as follows. Let $B=\exp(-2 \delta) A$. Then, from
(17) and (18) we obtain
\begin{equation}
  B'= \alpha r \left( A F'^2 - \frac{\sin^2{F}}{r^2}\right)
  e^{-2\delta}
\end{equation}
Using the boundary conditions (19), this implies that  $B(r)
\equiv 1$, hence $A = \exp(2\delta)$. Substituting this into (16)
we get
\begin{equation}
\frac{1}{r} e^{\delta} \left( r e^{\delta}  F'\right)' =
\frac{\sin(2 F)}{2 r^2}.
\end{equation}
Using the new coordinate $\rho$ defined by
\begin{equation}
  r e^{\delta} \frac{d}{dr} = \rho \frac{d}{d\rho},
\end{equation}
one can rewrite equation (21) in the form of the flat space
equation (12)
\begin{equation}
  \frac{1}{\rho} \frac{d}{d\rho} \left(\rho
  \frac{dF}{d\rho}\right) = \frac{\sin(2 F)}{2\rho^2},
\end{equation}
which, as we showed above, is solved by $F_S(\rho)=2
\arctan(\rho/\lambda)$ .
 Inserting this
solution into equation (17) and integrating  we get
\begin{equation}
  e^{\delta} = 1 - \frac{2\alpha \rho^2}{\lambda^2+\rho^2}.
\end{equation}
This yields the metric
\begin{equation}
  ds^2 = - dt^2 + (\lambda^2+\rho^2)^{-2\alpha} (d\rho^2 + \rho^2
  d\phi^2)
\end{equation}
which has the deficit angle equal to $4 \alpha \pi$, hence this
solution exists only for $\alpha<1/2$.
\section{Non-existence of static solutions for $\Lambda<0$}
For a nonvanishing cosmological constant, the static equations
(16) and (17) do not change while equation (18) picks up an
additional term
\begin{equation}
 A' = - 2 \Lambda r - \alpha r \left( A F'^2 +
 \frac{\sin^2{\!F}}{r^2}\right).
\end{equation}
Assuming that $\Lambda<0$, by rescaling, without loss of
generality, we set hereafter $\Lambda=-1$. Using (17) we eliminate
$\delta$ from (16) and get the following system
\begin{equation}
 A' =  2  r - \alpha r \left( A F'^2 +
 \frac{\sin^2{F}}{r^2},\right),
\end{equation}
\begin{equation}
  F'' + \frac{1}{r} F'  + \frac{2 r^2 - \alpha \sin^2{F}}{A r} F'
  = \frac{\sin(2 F)}{2 A r^2}.
\end{equation}
The boundary conditions at $r=0$ are
\begin{equation}
  F(r)\sim b r, \qquad A(r) \sim 1-(\alpha b^2-1) r^2.
\end{equation}
We want a solution for which $A \sim r^2$ at infinity and
$F(\infty)=\pi$. Such a solution was claimed to exist and
constructed numerically in \cite{km}.  We shall show now that this
claim was erroneous.

Let us define a function
\begin{equation}\label{H}
    H = \cos^2{F} + r^2 A F'^2.
\end{equation}
We claim that $H(r)$ is monotone decreasing. To prove this, using
the field equations, we compute
\begin{equation}\label{H'}
    H' = -2 r^3 F'^2 - \alpha r^3 A F'^4 + \alpha r \sin^2{\!F} F'^2.
\end{equation}
It is convenient to rewrite equation (\ref{H'}) in the form
\begin{equation}\label{G}
H' = - r F'^2 G(r), \quad G(r)= 2 r^2- \alpha r^2 A F'^2 + \alpha
\sin^2{\!F}.
\end{equation}
From the boundary conditions (29) $G(r) \sim 2 r^2>0$ near $r=0$,
independent of $b$ and $\alpha$. Now we shall show that $G(r)=0$
implies $G'(r)\geq 0$. To this end we compute
\begin{equation}\label{G'}
    G'(r)= -2 \alpha r^3 F'^2 - \alpha^2 r^3 A F'^4 +4 r +
    \alpha^2 r \sin^2{F} F'^2.
\end{equation}
To evaluate $G'$ when $G=0$ we solve $G=0$ for $\sin^2{F}$ and
substitute that value into equation (\ref{G'}). We get
\begin{equation}\label{G'2}
    G'\Bigr\rvert_{G=0}= 4 r.
\end{equation}
Thus $G'(r)>0$ for $r>0$, and therefore $H'(r)\leq 0$ for all $r$.
Since $H(0)=1$, this implies that $H(r)<1$ for all $r>0$. This
excludes existence of a solution
  having $\lim_{r\rightarrow \infty}
F(r)=\pi$ because that would mean $\lim_{r\rightarrow \infty} H(r)
\geq 1$.

  In view of Struwe's result mentioned above, the nonexistence of a static
  nonconstant soliton suggests (but by no means proves) that the
  negative cosmological constant might act as a cosmic censor in
  this model.

  \section*{Acknowledgement}
  PB thanks Lars Andersson for discussions.
The research of PB was supported in part by the KBN grant 2 P03B
006 23 and  the FWF grant P15738.

\end{document}